\documentclass{svproc}
\usepackage{url}

\usepackage{graphicx}
\usepackage{color}

\newcommand{\be}{\begin{eqnarray}}
\newcommand{\ee}{\end{eqnarray}}
\newcommand{\bfp}{{\bf p}_{\perp}}  
\newcommand{\bfP}{{\bf P}} 

\begin{document}
\mainmatter              
\title{Parton distribution functions of proton in a light-front quark-diquark model}
\titlerunning{PDFs of proton in LFQDM}  
%
\author{Tanmay Maji \and Dipankar Chakrabarti}
%
%
%
\institute{Department of Physics, Indian Institute of Technology Kanpur, Kanpur 208016, India\\
\email{tanmay@iitk.ac.in, dipankar@iitk.ac.in}}
\maketitle              
\vspace*{-\baselineskip}
\begin{abstract}
We present the parton distribution functions(PDFs) for unpolarised, longitudinally polarized and transversely polarized quarks in a proton using the light-front quark diquark model. 
We also present the scale evolution of PDFs and calculate axial charge and tecsor charge for $u$ and $d$ quarks at a scale of experimental findings.
\end{abstract}
%
%
\vspace*{-\baselineskip}
In recent years, there have been a lot of activities to investigate the polarised PDFs and three dimensional structure of nucleons. Different model investigations gave many interesting insights into the nucleon structure\cite{models_ref1,models_ref2}. Light front AdS/QCD predicts a particular form of wave function for a two body bound state\cite{BT} which can predict many interesting nucleon properties in quark-diquark model. Light front AdS/QCD wave function also encodes the non-perturbative ingredients of nucleons. Recently, the light front wave functions(LFWFs) for the nucleons in  quark-scalar-diquark models\cite{CM1,valery} have been constructed from the light-front AdS/QCD prediction.
In \cite{MC_model}, the contribution of vector diquark is included considering the spin-flavour SU(4) structure and different properties of nucleons are studied in detail. In this paper, we present the important predictions of the model of \cite{MC_model} and compared with available data.
\vspace*{-\baselineskip}
\section*{PDFs in Light-front quark-diquark model}
%
In the quark-diquark model, the nucleon is treated as a bound state of a quark and a diquark. The proton state is written in the spin-flavour SU(4) structure as 
\be 
|P; \pm\rangle = C_S|u~ S^0\rangle^\pm + C_V|u~ A^0\rangle^\pm + C_{VV}|d~ A^1\rangle^\pm. \label{PS_state}
\ee 
Where $|u~ S^0\rangle,~|u~ A^0\rangle$ and $|u~ A^0\rangle$ are two particle state with isoscalar-scalar, isoscalar-axial vector and isovector-axial vector diquarks  respectvely\cite{models_ref1}. 
The states are written by the two particle Fock state expansion with $J^z =\pm1/2$ for both the scalar and the axial vector diquarks.
We adopt the LFWFs $\varphi^{(\nu)}_i(x,\bfp)$ from soft-wall AdS/QCD prediction\cite{BT} and modify as
\be
\varphi_i^{(\nu)}(x,\bfp)=\frac{4\pi}{\kappa}\sqrt{\frac{\log(1/x)}{1-x}}x^{a_i^\nu}(1-x)^{b_i^\nu}\exp\bigg[-\delta^\nu\frac{\bfp^2}{2\kappa^2}\frac{\log(1/x)}{(1-x)^2}\bigg].
\label{LFWF_phi} 
\ee
The wave functions $\varphi_i^\nu ~(i=1,2)$ reduce to the AdS/QCD prediction for the parameters $a_i^\nu=b_i^\nu=0$  and $\delta^\nu=1.0$.
 We use the AdS/QCD scale parameter $\kappa =0.4~GeV$ as determined in \cite{CM1} and the quarks are  assumed  to be  massless. The parameters are fixed by fitting the Dirac and Pauli form factors. More detail of this formalism is given in \cite{MC_model}.
%
%

The parton distributions encodes the probability of finding a quark having light-cone momentum fraction $x=p^+/P^+$ in a proton. The PDFs are defined,at equal ligt-front time $z^+=0$, as 
\be 
\Phi^{\Gamma(\nu)}(x)=\frac{1}{2}\int \frac{d z^-}{2(2\pi)} e^{ip^+z^-/2} \langle P;S|\bar{\psi}^{(\nu)}(0)\Gamma\mathcal{W}_{[0,z]}\psi^{(\nu)}(z^-)|P; S\rangle \bigg|_{z^+=z_T=0},
\ee
where different  Dirac structures e.g., $\Gamma=\gamma^+, \gamma^+ \gamma^5$ and $ i\sigma^{j+}\gamma^5$ gives unpolarized PDF $f_1(x)$, helicity distribution $g_1(x)$ and transversity distribution $h_1(x)$ respectively.
$|P; S\rangle$ is the proton state  having spin $S$ and momentum $\bfP$. The quark fields $\psi$ and $\overline{\psi}$, at two different points $z^-$ and $0$, are connected by a Wilson line $\mathcal{W}_{[0,z]}$. At the light-cone gauge $A^+=0$ the Wilson line becomes unity. 
\begin{figure}[htbp]\includegraphics[width=4.cm,height=3.cm]{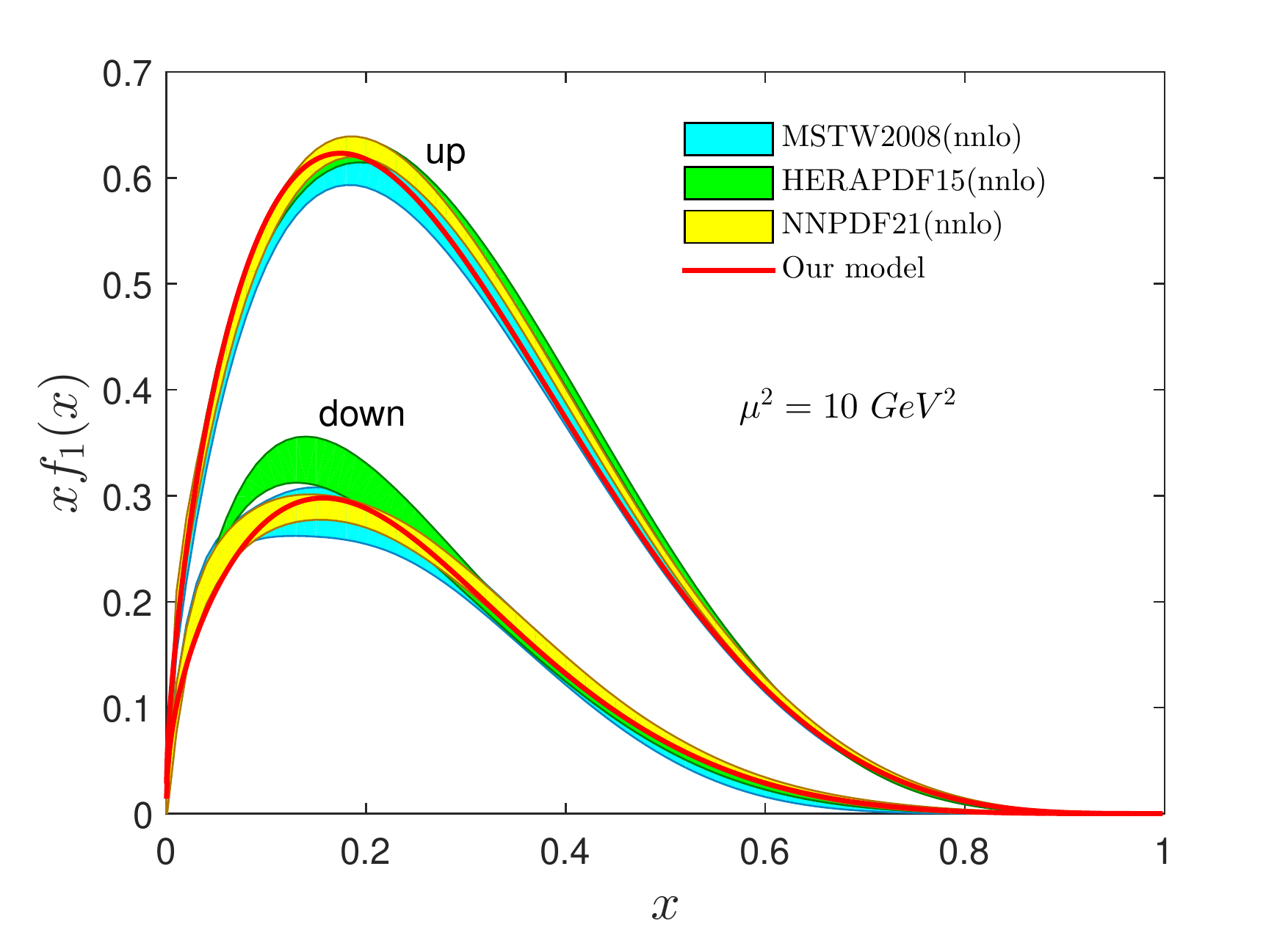}
\includegraphics[width=4.cm,height=3.cm]{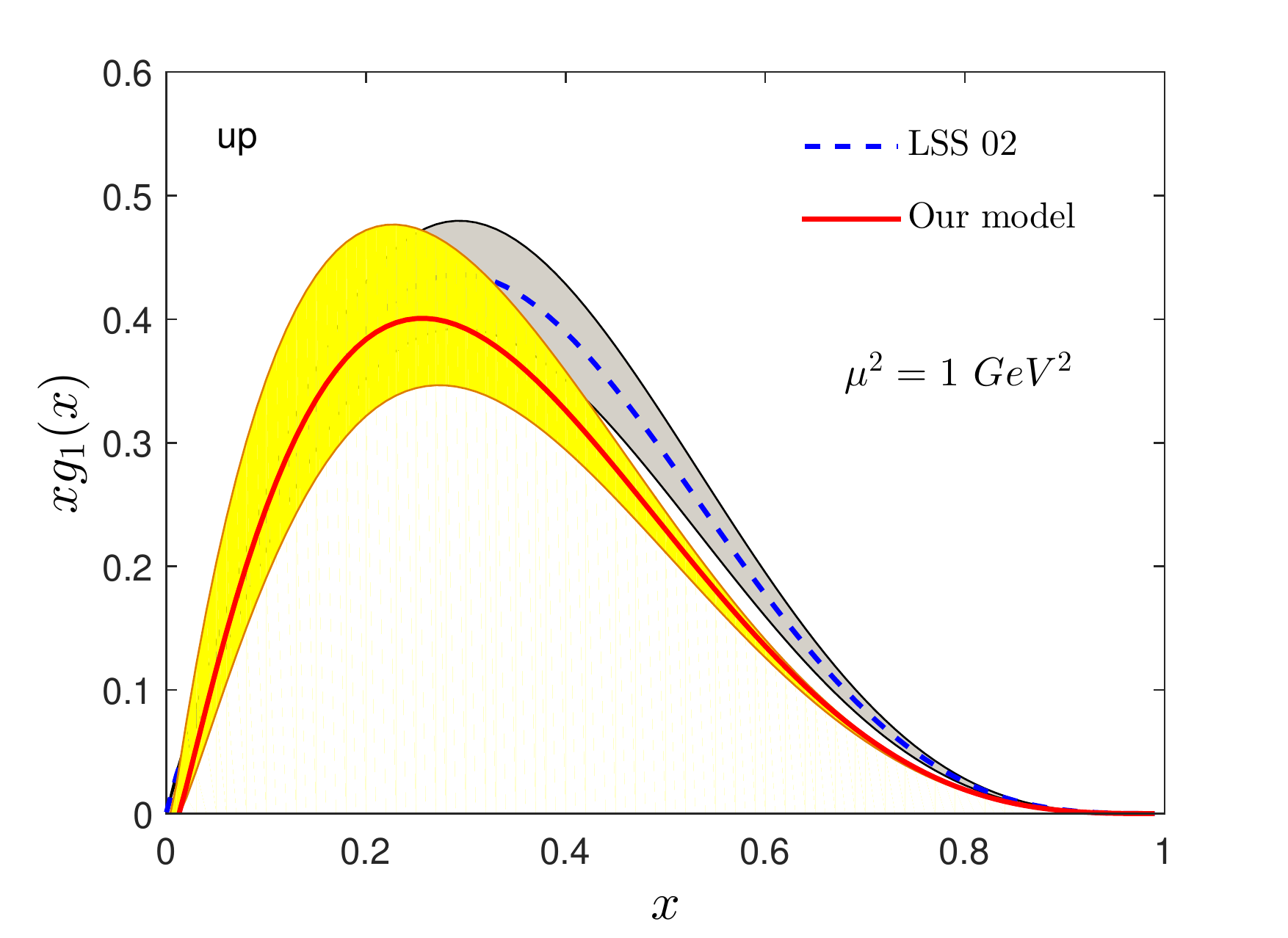}
\includegraphics[width=4.cm,height=3.cm]{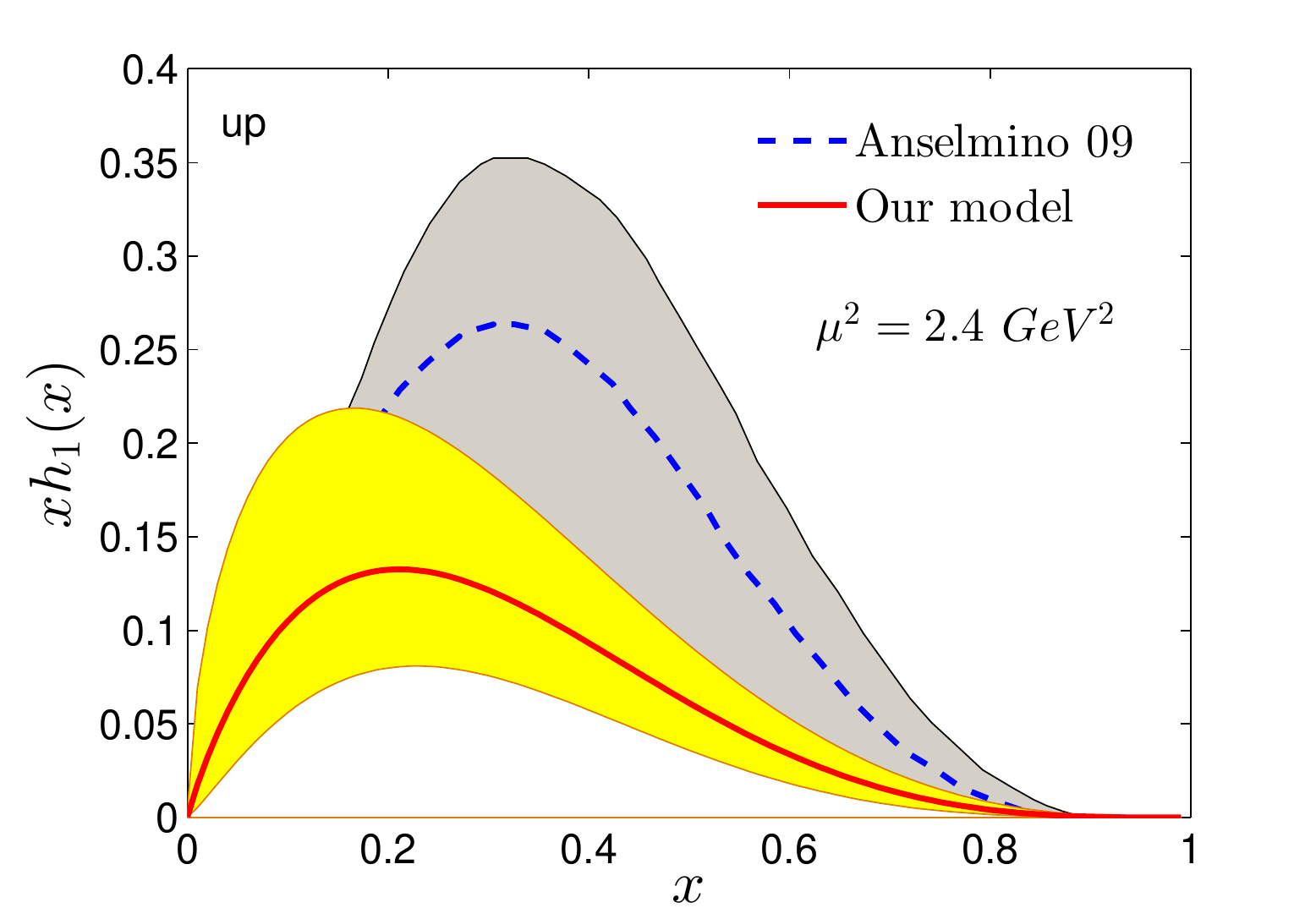}\\
\includegraphics[width=4.cm,height=3.cm]{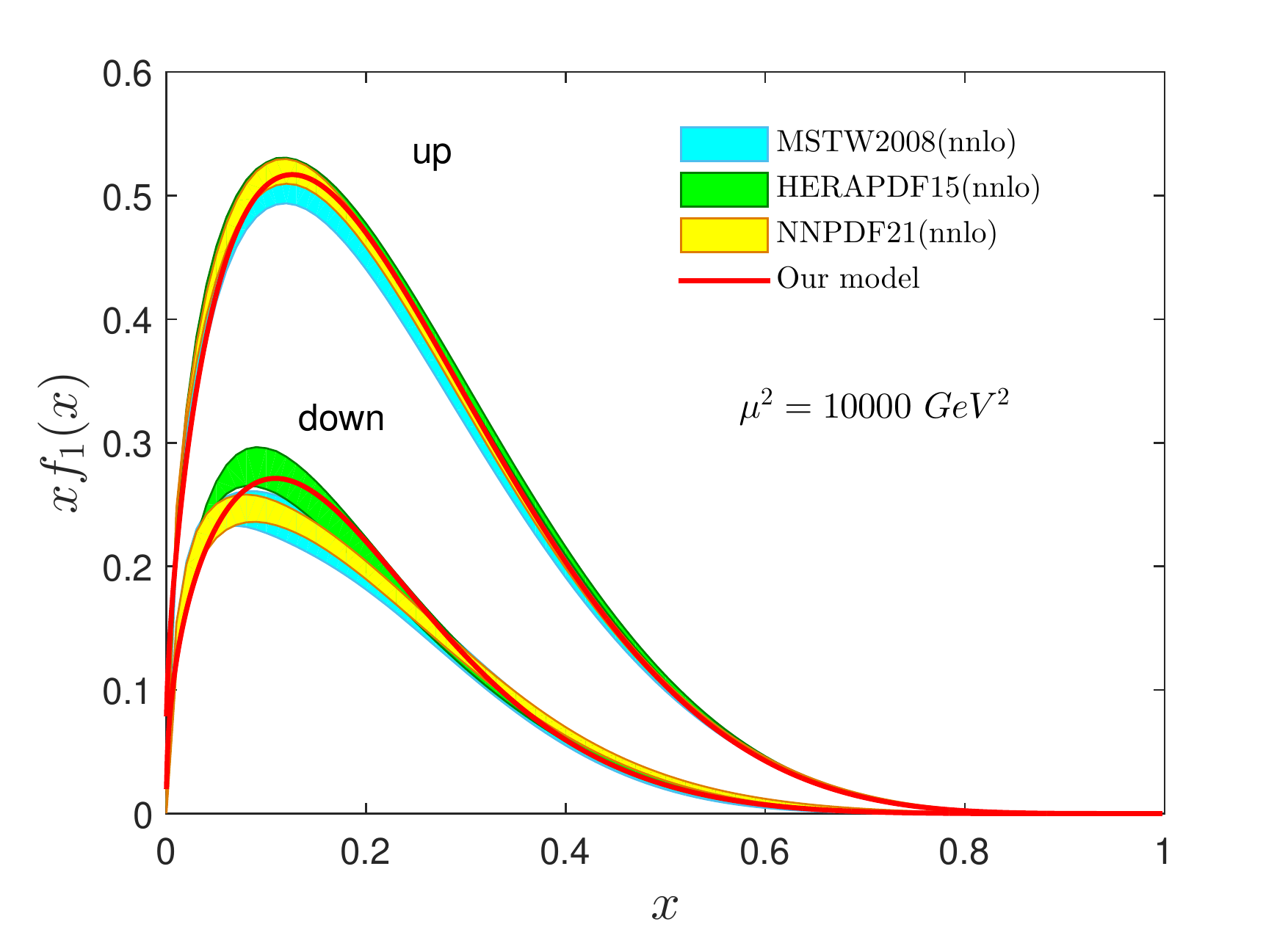}
\includegraphics[width=4.cm,height=3.cm]{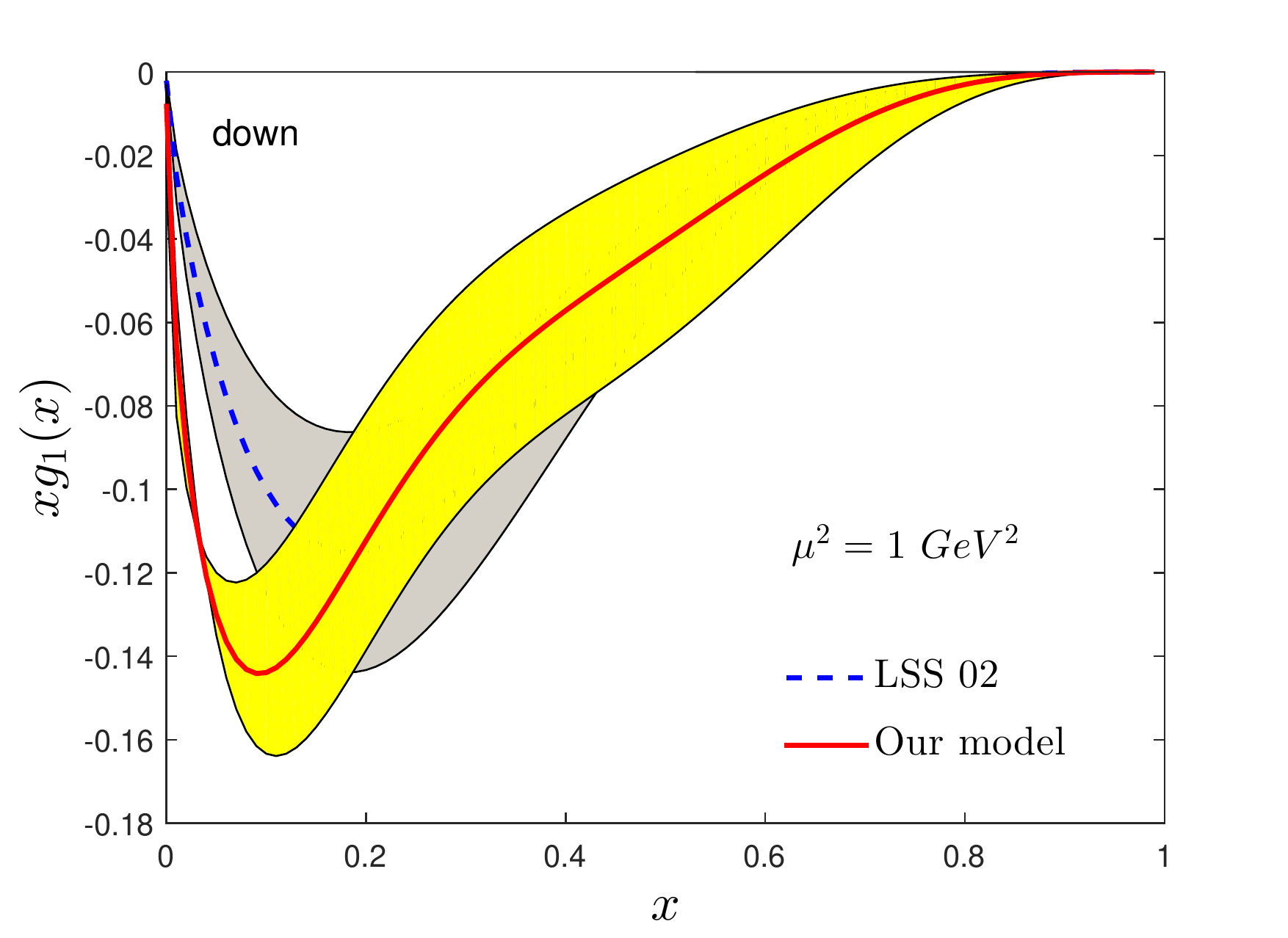}
\includegraphics[width=4.cm,height=3.cm]{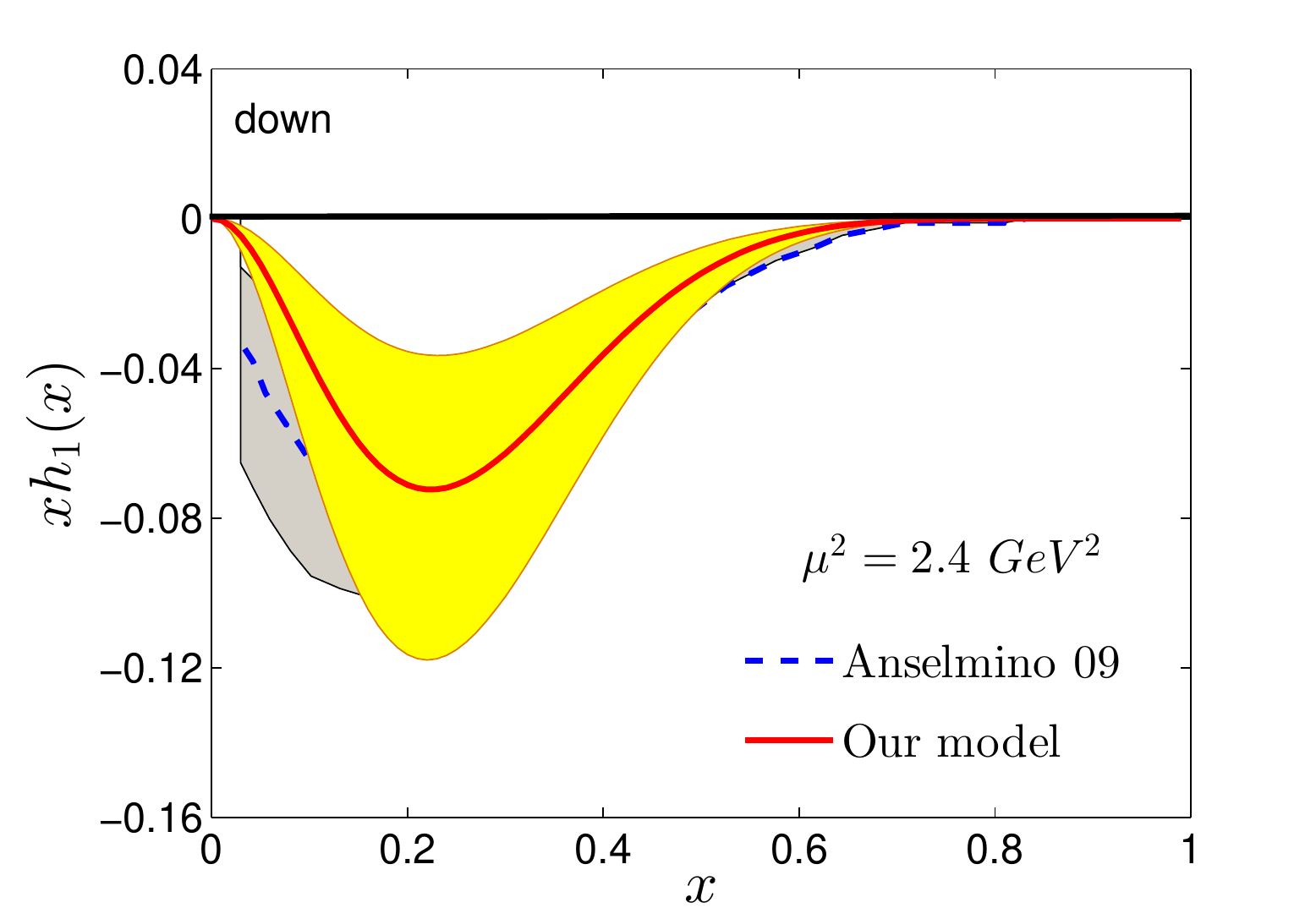}
\caption{\label{fig_PDF} Model prediction of $x f_1(x),~ x g_1(x)$ and $ x h_1(x)$ are shown by the continuous lines(red) for both $u$ and $d$ quarks. Left column: unpolarized PDF at  $\mu^2=10$ and $10000$ GeV$^2$  and compared with experimental data and the phenomenological extractions \cite{HERAPDF,NNPDF,MSTW}. Middle column: helicity distribution at $\mu^2=1~ GeV^2$. Last column: transversity distribution at $\mu^2=2.4~ GeV^2$.} 
\end{figure}

Since the different experiments have different energy scales, scale evolutions of PDFs are needed to compare with the experimental data. We simulate the DGLAP evolution of the PDF  by making the parameters in the unpolarized PDF  scale dependent where the values of the parameters at $\mu_0$ are  the same as in the LFWFs. The scale evolution in this model is discussed in \cite{MC_model}. The explicit expression of unpolarised PDFs are given as
\be 
f^{\nu}_1(x,\mu)&=& N^\nu_{f_1}(\mu) \bigg[\frac{1}{\delta^\nu(\mu)} x^{2a_1^\nu(\mu)}(1-x)^{2b_1^\nu(\mu)+1}\nonumber +\frac{\kappa^2 x^{2a_2^\nu(\mu)-2}(1-x)^{2b_2^\nu(\mu)+3}}{(\delta^\nu(\mu))^2M^2\ln(1/x)}\bigg],
\ee
with a flavour $\nu$ and scale $\mu$. $N^\nu_{f_1}(\mu)$ is the normalization constant.
The model prediction for unpolarized PDFs, at the scale $\mu^2=10$ and $10^4~GeV^2$, are shown in Fig.\ref{fig_PDF}(left column) for $u$ and $d$ quarks. We compare our result with the experimental data and phenomenological extractions e.g., HERAPDF15(nnlo), NNPDF21(nnlo) and MSTW2008(nnlo). The PDFs are measured by the H1 and the ZEUS collaborations from the QCD analysis in exclusive $e^\pm p$ scattering at the HERA\cite{HERAPDF}. The NNPDF provides unpolarised PDFs at the next-to-next leading order(NNLO) using the Neural Network approach to the deep-inelastic data\cite{NNPDF}. The MSTW2008 proposed phenomenological PDFs from the global fit analysis of hadron scattering data at LHC\cite{MSTW}.

Helicity PDF $g_1(x)$ provides the information about longitudinal polarized quark. The model prediction for the helicity PDFs are shown in Fig.\ref{fig_PDF}(middle column) at $\mu^2=1~GeV^2$. The error bands in this model(yellow) come from uncertainties in the model parameters and error bands in phenomenological extractions(gray) are from a constant relative error of $10\%$ to $g^u_1$ and $25\%$ to $g^d_1$ in the data taken from\cite{LSS02}. The axial charges, $g_A=g_A^u-g_A^d$, which are obtained from the first moment of the helicity distributions are given in Table.\ref{tab} and compared with the measured data\cite{Lead10}. The model predictions agree with the experimental data within the error bars.
\begin{table}[ht]
\centering 
\begin{tabular}{|c|c|c|c||c|c|c|}
 \hline
 & $g^u_A$& $g^d_A$&~$g_A$ & $g^u_T$~~&~~ $g^d_T$~&~$g_T$\\ \hline
 LFQDM & $0.71\pm0.09$ & $-0.54^{+0.19}_{-0.13}$ & $1.25^{+0.28}_{-0.22}$ & $0.37^{+0.06}_{-0.05}$ ~&~ $-0.14^{+0.05}_{-0.06}$~&~ $0.51^{+0.12}_{-0.11}$ \\ 
  Measurement & $0.82\pm 0.07$ & $-0.45\pm 0.07$ & $1.27\pm0.14$ & $0.59^{+0.14}_{-0.13}$ ~&~ $-0.20^{+0.05}_{-0.07}$~&~ $0.79^{+0.19}_{-0.20}$ \\
\hline
 \end{tabular} 
\caption{Axial charge $g^\nu_A$ at $\mu^2=1~ GeV^2$ and tensor charge $g^\nu_T$ at $\mu^2=0.8~ GeV^2$ are listed and compared with the mesaurements\cite{Lead10,Anse09}.} 
\label{tab} 
\end{table}
The probability of finding a transversely polarized quark is given by the transversity PDF $h_1(x)$. The model predictions are shown to agree with the experimental data\cite{Anse09}, see Fig.\ref{fig_PDF}(last column), at $\mu^2=2.4~GeV^2$ for both $u$ and $d$ quarks. The first moment of the transversity distribution gives the tensor charge denoted by $g_T$. The model again predicts the tensor charges quite accurately as shown in Table.\ref{tab} and fall within the uncertainty bands of the phenomenological fits for both $u$ and $d$ quark\cite{Anse09}. We observe $\mid g_T^\nu\mid  < \mid g_A^\nu\mid$. The scale independent ratio $\mid {g_T^d}/ {g_T^u}\mid =0.38$ is very close to the phenomenological prediction\cite{Anse09}. This model also satisfies the Soffer bound $|h^\nu_1(x,\mu)|\leq \frac{1}{2}\big[ f^\nu_1(x,\mu)+g^\nu_1(x,\mu) \big]$ at an arbitrary scale \cite{Soff95}.
%
%

We show the prediction of the unpolarised and polarised PDFs in the quark-diquark model where the wave function is constructed from the soft-wall AdS/QCD prediction. Light-front AdS/QCD can  accurately predict the PDF evolution up to a very high scale ($\mu^2=10^4~ GeV^2$). The helicity and transversity PDFs are calculated as predictions of the model and  have good agreement with the available data. The transversity distribution satisfies the Soffer bound. Our model reproduces the experimental values of axial and tensor charges quite well. 
%
%

\end{document}